\documentclass[12pt]{article}

\usepackage{amsfonts}
\usepackage{amssymb}
\usepackage{amsbsy}
\usepackage{amsmath}
\usepackage{epsf}

\newcommand{\BEQ}{\begin{equation}}     
\newcommand{\BEA}{\begin{eqnarray}}
\newcommand{\be}{\begin{equation}}
\newcommand{\bel}[1]{\begin{equation}\label{#1}}
\newcommand{\EEQ}{\end{equation}}
\newcommand{\EEA}{\end{eqnarray}}
\newcommand{\ee}{\end{equation}}

\newcommand{\smfrac}[2]{\mbox{\small$\frac{#1}{#2}$}}
\newcommand{\eps}{\varepsilon}          
\newcommand{\vph}{\varphi}              
\newcommand{\D}{{\rm d}}                
\newcommand{\II}{{\rm i}}               
\newcommand{\wit}[1]{\widetilde{#1}}    
\renewcommand{\vec}[1]{{\boldsymbol{#1}}} 

\def\bbbz{{\mathchoice {\hbox{$\sf\textstyle Z\kern-0.4em Z$}}
{\hbox{$\sf\textstyle Z\kern-0.4em Z$}}
{\hbox{$\sf\scriptstyle Z\kern-0.3em Z$}}
{\hbox{$\sf\scriptscriptstyle Z\kern-0.2em Z$}}}}

\catcode`\@=11
\def\numberbysection{\@addtoreset{equation}{section}
  \def\theequation{\thesection.\arabic{equation}}}
\numberbysection

\begin{document}
\begin{titlepage}

\medskip

\begin{center}
{\Large \bf On nonlinear partial differential equations with an
infinite-dimensional conditional symmetry}
\end{center}

\centerline{{\bf Roman Cherniha}$^{a,b}$\footnote{\small
e-mail:  {\tt cherniha@imath.kiev.ua}} and
{\bf Malte Henkel}$^b$\footnote{\small e-mail: {\tt henkel@lpm.u-nancy.fr} }}
\vskip 0.5 cm
\centerline{$^a$Institute of Mathematics, National Academy of Science
of Ukraine}
\centerline{3, Tereshchenkivs'ka Str., UA - 01601 Kyiv, Ukraine}
\centerline{$^b$Laboratoire de Physique des Mat\'eriaux,\footnote{Laboratoire
associ\'e au CNRS UMR 7556} Universit\'e Henri Poincar\'e Nancy I,}
\centerline{B.P. 239, F -- 54506 Vand{\oe}uvre l\`es Nancy Cedex, France}


\begin{abstract}
The invariance of  nonlinear partial differential equations under
a certain infinite-dimensional Lie algebra $\mathfrak{A}_N(z)$ in
$N$ spatial dimensions is studied. The special case
$\mathfrak{A}_1(2)$ was introduced in J. Stat. Phys. {\bf 75},
1023 (1994) and contains the Schr\"odinger Lie algebra
$\mathfrak{sch}_1$ as a Lie subalgebra. It is shown that there is
no second-order equation which is invariant under the massless
realizations of $\mathfrak{A}_N(z)$. However, a large class of
strongly non-linear partial differential equations is found which
are conditionally invariant with respect to the massless
realization of $\mathfrak{A}_N(z)$ such that the well-known
Monge-Amp\`ere equation is the required additional condition. New
exact solutions are found for some representatives of this class.
\end{abstract}

\end{titlepage}

\section{Introduction}

The maximal invariance algebra (MIA) of the free Schr\"odinger
equation in $N$ spatial dimensions has been studied since a long
time. This Lie algebra is known as the {\em Schr\"odinger algebra}
\cite{Nied72} and will be denoted here by $\mathfrak{sch}_N$. An
infinite-dimensional generalization of $\mathfrak{sch}_1$ was
constructed in \cite{Henk94} and will be denoted here by
$\mathfrak{A}_1(2)$. These algebras contain infinitesimal
dilatations in `time' and `space' coordinates, respectively, and
correspond to the finite transformations $t\mapsto b^2 t$,
$\vec{x}\mapsto b \vec{x}$ where $b$ is an arbitrary real
parameter and $\vec{x}=(x_1,\ldots,x_N)$. While the
transformations contained in $\mathfrak{sch}_N$ merely contain the
projective transformation $t\mapsto (\alpha t+\beta)/(\gamma t
+\delta)$ with real parameters satisfying the restriction
$\alpha\delta-\beta\gamma=1$, the transformations in
$\mathfrak{A}_N(2)$ allow for {\it arbitrary} transformation in
`time', viz. $t\mapsto b(t)^2$, where $b$ is an arbitrary smooth
function. Recently, this construction was extended to an arbitrary
{\em dynamical exponent} $z$ defined by the dilatation
\BEQ
t\mapsto t' := b^z t \;\; , \;\; \vec{x} \mapsto \vec{x}' := b \vec{x}
\EEQ
An infinite-dimensional Lie algebra
$\mathfrak{A}_N(z)$ can be obtained by allowing transformation in
`time' $t\mapsto b(t)^z$ \cite{Henk02}. In order to define it,
consider the generators
\BEQ \label{1:gl:X0}
X_n = z t^{n+1}\partial_t +(n+1)t^n {x}_a\partial_a
      + \lambda (n+1)t^n u\partial_u
\EEQ
and
\BEA
Y_m^{(a)} &=& t^{m+1/z}\partial_a  \label{1:gl:XYJ}
\\
J_{ab} &=& x_a\partial_b - x_b\partial_a \;\;\;\; ; \;\; a,b=1,\ldots,N
\nonumber
\EEA
where the summation convention over repeated indices is here and afterwards
implied, $\partial_a=\partial/\partial {x}_a$, $\partial_t=\partial/\partial t$,
$\partial_u=\partial/\partial u$ (where $u$ is the function the generators are
supposed to act on). Finally $n\in\mathbb{Z}$ and $m+\frac{1}{z}\in\mathbb{Z}$
will be used throughout.
The non-vanishing commutators of these generators are \cite{Henk02}
\BEQ \label{1:gl:Komm}
\left[ X_n, X_{n'}\right] = z\left(n'-n\right) X_{n+n'} \;\; , \;\;
\left[ X_n, Y_m^{(a)}\right] = \left( z m-n\right) Y_{n+m}^{(a)} \;\; , \;\;
\left[ Y_m^{(a)}, J_{bc}\right] = \delta_{ab} Y_m^{(c)} - \delta_{ac} Y_m^{(b)}
\EEQ

In this paper we shall be concerned with two infinite-dimensional
Lie algebras which are defined in terms of a basis of generators
from (\ref{1:gl:X0}) and (\ref{1:gl:XYJ}) as follows.
\BEA
\mathfrak{A}_N(z) &:=& \left\{ X_n, Y_m^{(a)} \left|
n\in\mathbb{Z},\: m+\frac{1}{z}\in\mathbb{Z},\: a\in\{1,\ldots,N\}
\right.\right\} \label{1:gl:A}
\\
\mathfrak{B}_N(z) &:=& \left\{ X_n, Y_m^{(a)} \left|
n\in\{-1,0\},\: m+\frac{1}{z}\in\mathbb{Z},\: a\in\{1,\ldots,N\}
\right.\right\} \label{1:gl:B}
\EEA
Obviously, $\mathfrak{B}_N(z)$
is a Lie subalgebra of $\mathfrak{A}_N(z)$ and $\mathfrak{sch}_1$
is the maximal finite-dimensional Lie subalgebra of
$\mathfrak{A}_1(2)$. Since with the definition (\ref{1:gl:XYJ})
the commutator $[Y_m^{(a)},Y_{m'}^{(b)}]=0$
for  all pairs $(m,m')$, the generators (\ref{1:gl:XYJ}) are
said to form a {\it massless} realization of the Lie algebras
$\mathfrak{A}_N(z)$ or $\mathfrak{B}_N(z)$, respectively. For the
massless realization (\ref{1:gl:XYJ}) only the arguments of the
function $u$ but not $u$ itself transform when acting with the
generators $Y_m^{(a)}$.

Eqs.~(\ref{1:gl:X0}) and (\ref{1:gl:XYJ}) can be extended to
massive realizations of $\mathfrak{A}_N(2)$ as given in
\cite{Henk94,Henk02} and its has been shown that these may be used
as a a dynamical symmetry in non-equilibrium statistical
mechanics, in particular in relation with ageing phenomena
\cite{Henk02,Henk03}. Massive realizations of the
infinite-dimensional Lie algebra $\mathfrak{A}_1(1)$ are also
known \cite{Henk02}.
For $z\ne 1,2$, massive realizations can no longer be constructed
in terms of first-order differential operators, see \cite{Henk02}
for details and for applications to physical ageing. On the other
hand, massless realizations may also be of physical interest. For
example, it is well-known that the $1D$ Burgers equation $u_t+u
u_x-u_{xx}=f(t,x,U)$ has a massless realization of
$\mathfrak{sch}_1$ (albeit different from (\ref{1:gl:X0}) and
(\ref{1:gl:XYJ})) for an external force $f=\mbox{\rm constant}$
(and moreover $f=0$ can be achieved by a non-trivial local
substitution) \cite{Ivas97,Cher98}.

We shall attempt to look for non-linear partial differential
equations (PDEs) which have the infinite-dimensional massless
realization (\ref{1:gl:X0}) and (\ref{1:gl:XYJ}) of
$\mathfrak{A}_N(z)$ or $\mathfrak{B}_N(z)$ as dynamical symmetry.

The so-called scaling-dimension $\lambda\ne 0$ can always be
brought to $\lambda=1$ by the local substitution $u\to
u^{\lambda}$. Therefore, without loss of generality, we may write
\BEQ \label{1:gl:X}
X_n = z t^{n+1}\partial_t +(n+1)t^n {x}_a\partial_a + (n+1)t^n u\partial_u
\EEQ
instead of (\ref{1:gl:X0}) and we
shall do so throughout this paper. Note that the Lie algebra
$\mathfrak{A}_N(z)$ contains the following  subalgebra spanned by
the generators
\BEQ \label{1:gl:Gal}
Y_{-1/z}^{(a)} = P_a =\partial_a \;\; , \;\;
Y_{1-1/z}^{(a)} = G_a^0 = t\partial_a \;\; , \;\;
X_{-1} = P_t =\partial_t
\EEQ
with $a,b=1,\ldots,N$. It is easy to see that these
generators together with $J_{ab}$ (see (\ref{1:gl:XYJ})) span the
Galilei algebra $AG^0(1,N)$ with a vanishing mass operator. The
partial differential equations invariant with respect to the
algebra $AG^0(1,N)$ and its subalgebra (\ref{1:gl:Gal}) were
classified in \cite{Fush85,Cher85,Cher01}.

Starting form this observation, we prove in section~2 that there
is no second-order PDE invariant under the Lie algebra
$\mathfrak{A}_N(z)$ but we do find PDEs with the Lie algebra
$\mathfrak{B}_N(z)$ as a dynamical symmetry, for any given value
of $z$. In addition, we construct a wide class of PDEs which admit
$\mathfrak{A}_N(z)$ as a conditional dynamical symmetry and we
show that the additional condition is the $N$-dimensional
Monge-Amp\`ere equation. Monge-Amp\`ere equations are classical
equations in differential geometry and analysis and arise in many
different contexts. For recent reviews of their mathematical
theory, see \cite{Caff99,Guti01,Schu90}. Among the physical
applications, we quote the discussion of the equivalence of the
mass-transport problem with a quadratic cost function to a
(inhomogeneous $3D$) Monge-Amp\`ere equation and applications to
cosmology in \cite{Bren03}. The two-dimensional Monge-Amp\`ere
equations also plays a central r\^ole arises in the description of
pattern formation in convective systems modeled through the
Cross-Newell equation \cite{Erco03,Newe96}, as well as in M-theory
extensions of quantum chromodynamics \cite{Volo99}.

In section~3, we use the conditional symmetry to derive
exact solutions for some of these invariant equations. In particular, we study
the system of (1+2)-dimensional non-linear equations
\BEA
\det\left[\begin{array}{ccc} u_t    & u_1    & u_2 \\
                             u_{t1} & u_{11} & u_{12} \\
                             u_{t2} & u_{21} & u_{22} \end{array}\right]
&=& \frac{\partial}{\partial x_1} \left( u^{-z} u_1\right) +
\frac{\partial}{\partial x_2} \left( u^{-z} u_2\right)
\label{1:gl:inveq} \\
u_{11}u_{22}-u_{12}^2 &=& 0
\label{1:gl:inveqMA}
\EEA
where $u_t=\partial_t u$, $u_a=\partial u/\partial x_a$ and
$u_{ab}=\partial^2 u/\partial x_a\partial x_b$ with $a,b=1,\ldots,N$. The
Monge-Amp\`ere equation associated to (\ref{1:gl:inveq}) is given
by (\ref{1:gl:inveqMA}). Eq.~(\ref{1:gl:inveq}) might be viewed as an
example of a PDE conditionally invariant under $\mathfrak{A}_2(z)$, but it
might also be thought of as non-linear diffusion equation where the
time-derivative $u_t$ has been generalized.
Our conclusions are presented in section~4.

\section{Invariance of a class of PDEs under the Lie algebras
$\mathfrak{A}_N(z)$ and $\mathfrak{B}_N(z)$}

Consider the following class of PDEs with first- and second-order derivatives
\BEQ \label{2:gl:PDE}
F\left(t,\vec{x},u,u_t,\underset{1}{u},u_{tt},
\underset{1}{u_t},\underset{11}{u}\right) =0
\EEQ
in $(1+N)$-dimensional time-space. Here and afterwards, we use the
notation $\vec{x}=(x_1,\ldots,x_N)$, $\underset{1}{u}=(u_1,\ldots,u_N)$,
$\underset{1}{u_t}=(u_{t1},\ldots,u_{tN})$ and
$\underset{11}{u}=(u_{11},\ldots,u_{1N},\ldots,u_{NN})$ and $F$ is an
arbitrary smooth function. Obviously, any PDE of the form (\ref{2:gl:PDE})
which admits $\mathfrak{A}_N(z)$ as given by the generators
(\ref{1:gl:XYJ},\ref{1:gl:X}) must also be invariant under the
Galilei subalgebra $AG^0(1,N)$ spanned by the generators (\ref{1:gl:Gal}).
The PDEs of the form (\ref{2:gl:PDE}) which admit $AG^0(1,N)$ as a
dynamical symmetry have been classified long ago. The following statement
holds true.

\noindent {\bf Lemma.} \cite[Theorem 11]{Fush85} {\it If  $W^I$
and $W^{II}$ are defined as follows}
\BEQ \label{2:gl:wI}
W^I = \det\left[ \begin{array}{cccc}
u_t    & u_1    & \cdots & u_N    \\
u_{t1} & u_{11} & \cdots & u_{1N} \\
\vdots & \vdots &        & \vdots \\
u_{tN} & u_{1N} & \cdots & u_{NN}
\end{array} \right]
\EEQ
\BEQ
W^{II} = \det\left[ \begin{array}{cccc}
u_{tt} & u_{t1} & \cdots & u_{tN} \\
u_{t1} & u_{11} & \cdots & u_{1N} \\
\vdots & \vdots &        & \vdots \\
u_{tN} & u_{1N} & \cdots & u_{NN}
\end{array} \right]
\EEQ
{\it then PDEs of the form}
\BEQ \label{2:gl:Pgal}
F_1\left(W^I, W^{II},u,\underset{1}{u},\underset{11}{u}\right) =0
\EEQ
{\it where $F_1$ is an arbitrary smooth function, are most general PDEs
of the form (\ref{2:gl:PDE}) which are invariant under the
Lie algebra spanned by the generators (\ref{1:gl:Gal}).}

In consequence, any PDE with $\mathfrak{A}_N(z)$ as a Lie symmetry
will be among the equations of the form (\ref{2:gl:Pgal}).
Furthermore, since the generators $Y_m^{(a)}$ of (\ref{1:gl:XYJ})
generate the following group continuous transformations
\BEQ \label{2:gl:Gruppe}
x_a' = x_a + v_a t^{m+1/z} \;\; , \;\;
t' =t \;\; , \;\;
u' = u
\EEQ
where $v_1,\ldots,v_N$ are real (or complex) group parameters it is
straightforward to check that $W^I,\underset{1}{u}$ and $\underset{11}{u}$ are
absolute differential invariants of the Lie group (\ref{2:gl:Gruppe}). On
the other hand, $W^{II}$ transforms non-trivially under the action
of transformations from (\ref{2:gl:Gruppe}) with $m\ne -1/z,
1-1/z$. Therefore, {\em any} PDE which is invariant under
the Lie algebra spanned by the generators
$\{Y_{m}^{(a)},X_{-1}=P_t\}$ as defined in (\ref{1:gl:XYJ})
must be of the form
\BEQ \label{2:gl:F2}
F_2\left( W^I,u,\underset{1}{u},\underset{11}{u}\right) =0
\EEQ
where $F_2$ is an arbitrary smooth function.

We now consider the continuous transformations generated from the
$X_n$. Their finite form is easily obtained from (\ref{1:gl:X})
\BEA
t' &=& \left|t^{-n} -z n\eps_n\right|^{-1/n}
= t\left|1-zn\eps_n t^n\right|^{-1/n}
\nonumber \\
x_a' &=& x_a \left|1-zn\eps_n t^n\right|^{-(n+1)/(zn)}
\label{2:gl:GX} \\
u' &=& u  \left|1-zn\eps_n t^n\right|^{-(n+1)/(zn)}
\nonumber
\EEA
where the $\eps_n$, with $n\in\mathbb{Z}$, are group parameters and
$a=1,\ldots,N$.

In view of the quite lengthy calculations which have to be
performed, it is useful to consider first the case $N=1$. In this
case equation (\ref{2:gl:F2}) takes the form
\BEQ \label{2:gl:f_1}
W^I =f\left(u,u_x,u_{xx}\right)
\EEQ
where
\BEQ \label{2:gl:wI_1}
W^I = \det\left[ \begin{array}{cc} u_t    & u_{x} \\
                                   u_{tx} & u_{xx}
\end{array}\right]
\EEQ
and $f$ is the inverse function of $F_2$ when solving for the
variable $W^I$. In addition, eq.~(\ref{2:gl:GX}) reduces to
\BEA
t' &=&  t\left|1-zn\eps_n t^n\right|^{-1/n}
\nonumber \\
x' &=& x \left|1-zn\eps_n t^n\right|^{-(n+1)/(zn)}
\label{2:gl:GX1} \\
u' &=& u  \left|1-zn\eps_n t^n\right|^{-(n+1)/(zn)}
\nonumber
\EEA
Assuming $z\ne 0$ and that the parameters $\eps_n$ are
sufficiently small, the absolute value in eqs.~(\ref{2:gl:GX1})
may be dropped and we find the following transformation laws of
the derivatives
\BEA
u_{x'}'   &=& u_x \nonumber \\
u_{x'x'}' &=& {A(t)}^{-1} u_{xx} \nonumber \\
u_{t'}'   &=& A(t)^{1-z} u_t + \left(u - xu_x\right) n(n+1)\eps_n t^{n-1}
A(t)^{1+zn/(n+1)}  \label{2:gl:TrAbl} \\
u_{t'x'}' &=& A(t)^{-z} u_{tx} - x u_{xx} n(n+1)\eps_n t^{n-1} A(t)^{zn/(n+1)}
\nonumber
\EEA
where
\BEQ A(t) := \left( 1 -zn\eps_n t^n\right)^{-(n+1)/(zn)}
\EEQ
Substituting the
transformation formulas (\ref{2:gl:GX1},\ref{2:gl:TrAbl}) into
eq.~(\ref{2:gl:f_1}) for the function $u'(t',x')$, we arrive at
the equation
\BEQ \label{2:gl:f}
A(t)^{-z} W^I + n(n+1)\eps_n t^{n-1} A(t)^{zn/(n+1)} u u_{xx}
= f\left( A(t) u, u_x, A(t)^{-1} u_{xx} \right)
\EEQ
Consider the function $B(t) := (n+1)\eps_n t^{n-1}$, $n\in\mathbb{Z}$.
It arises only in the second term on
the left-hand side of eq.~(\ref{2:gl:f}) and cannot be expressed
as some power of $A(t)$. Therefore, eq.~(\ref{2:gl:f}) is {\em
not} reducible to eq.~(\ref{2:gl:f_1}) for any function $f$ (and
not even in the special case $f=0$). The only exceptions to this
observations occur for $n=-1$ and $n=0$ when the second term on
the left-hand side vanishes. The operators $X_{-1}=P_t$ and
$X_0=D=zt\partial_t+x\partial_x+u\partial_u$ correspond to these cases.

It is easily checked that the same result is also obtained in the
case when $\left|1-zn\eps_n t^n\right|=zn\eps_n t^n-1$ which
arises when $\eps_n$ is not small.

Summarizing, we have just seen that there is no PDE
belonging to the class (\ref{2:gl:f_1}) which admits the
infinite-dimensional Lie algebra $\mathfrak{A}_1(z)$ with
generators given by eqs.~(\ref{1:gl:XYJ},\ref{1:gl:X}) and $z\ne 0$
as a dynamical symmetry.
On the other hand, we can look for those PDEs which are
invariant under the infinite-dimensional Lie algebra
$\mathfrak{B}_1(z)$.
In this case eq.~(\ref{2:gl:f}) simplifies into
\BEQ
A(t)^{-z} W^I = f\left( A(t) u, u_x, A(t)^{-1} u_{xx} \right)
\EEQ
and we see that this can be reduced to (\ref{2:gl:f_1}) only if
\BEQ \label{2:gl:g}
f = u^{-z} g\left( u_x, u u_{xx}\right)
\EEQ
where $g$ is an arbitrary smooth function of two variables.
Thus, we have proved the following theorem.

\noindent {\bf Theorem 1.} {\it (i) Any PDE belonging to the class
(\ref{2:gl:f_1}) cannot be invariant under the Lie algebra
$\mathfrak{A}_1(z)$ generated by (\ref{1:gl:XYJ},\ref{1:gl:X})
with $N=1$ and $z\ne 0$. \\ (ii) Only non-linear PDEs of the form}
\BEQ W^I = u^{-z} g\left(u_x,u u_{xx}\right) \EEQ {\it where $W^I$
is defined by (\ref{2:gl:wI_1}) are invariant under the
infinite-dimensional Lie algebra $\mathfrak{B}_1(z)$
spanned by the basis of operators}
\BEQ
X_{-1} = P_t =\partial_t \;\; , \;\;
X_{0} = D = zt\partial_t + x\partial_x + u\partial_u \;\; , \;\;
Y_m = t^{m+1/z} \partial_x \;\; ; \;\;
m+\frac{1}{z}\in \mathbb{Z}
\EEQ

\noindent {\bf Remark 1.} Consider the generators $X_n$ in the
case $z=0$. Then we have the continuous transformations
\begin{displaymath}
t'=t \;\; , \;\; x'=x\exp(\eps_n t^n) \;\; , \;\; u' = u\exp(\eps_n t^n)
\end{displaymath}
with $n\in\mathbb{Z}$. Simultaneously, we  replace the series of
the operators $ Y_m^{(a)}$ arising in (\ref{1:gl:XYJ}) by the
operators $\wit{Y}_{m}^{(a)}:=t^{m} \partial_a$ with $m\in\mathbb{Z}$.
Then it is easy to check that the result of Theorem 1 remains valid also for
the case $z=0$.

We now treat the multidimensional case $N>1$. It turns out that
the transformations (\ref{2:gl:GX}) lead to a simple
generalization of the one-dimensional formulas
Eqs.~(\ref{2:gl:TrAbl},\ref{2:gl:f},\ref{2:gl:g}). Indeed,
Eqs.~(\ref{2:gl:TrAbl}) now take the form
\BEA
u_{a'}'   &=& u_a \nonumber \\
u_{a'b'}' &=& A(t)^{-1} u_{ab} \nonumber \\
u_{t'}'   &=& A(t)^{1-z} u_t + \left( u-x_a u_a\right) n(n+1)\eps_nt^{n-1}
A(t)^{1+zn/(n+1)} \\
u_{t'b'}' &=& A(t)^{-z} u_{tb} - x_au_{ab}n(n+1)\eps_nt^{n-1}
A(t)^{zn/(n+1)} \nonumber
\EEA
Substituting this into a multidimensional analogue of
(\ref{2:gl:f}) we arrive at the expression
\BEQ \label{2:gl:Cond}
A(t)^{1-z-N} W^I + n(n+1)\eps_n t^{n-1} A(t)^{zn/(n+1)+1-N} u
W_N^{II} = f\left( A(t) u, \underset{1}{u}, A(t)^{-1}
\underset{11}{u}\right)
\EEQ
where
\BEQ \label{2:gl:wIII}
W_N^{II} = \det\left[\begin{array}{cccc}
  u_{11} & u_{12} & \cdots & u_{1N} \\
  u_{21} & u_{22} & \cdots & u_{2N} \\
  \vdots & \vdots &        & \vdots \\
  u_{N1} & u_{N2} & \cdots & u_{NN}
\end{array} \right]
\EEQ
We can now analyze Eq.~(\ref{2:gl:Cond}) along the same lines
as before (and also its analogue for the case $z=0$) and arrive at
the following result which generalizes Theorem 1.

\noindent {\bf Theorem 2.} {\it (i) Any partial differential
equation of the class (\ref{2:gl:PDE}) cannot be invariant under the Lie
algebra $\mathfrak{A}_N(z)$ generated by (\ref{1:gl:XYJ},\ref{1:gl:X}). \\
(ii) Equations of the form}
\BEQ \label{2:gl:Th2}
W^I = u^{1-z-N} g\left(\underset{1}{u}, u \underset{11}{u}\right)
\EEQ
{\it where $g$ is an arbitrary smooth
function of $N(N+3)/2$ variables, are the only PDEs of the class
(\ref{2:gl:PDE}) invariant under the infinite-dimensional Lie
algebra $\mathfrak{B}_N(z)$ (\ref{1:gl:B}).}

\noindent {\bf Remark 2.} The class (\ref{2:gl:Th2}) of PDEs
contains some equations belonging to the class of
reaction-diffusion-convection equations which were recently
suggested and investigated in \cite{Cher01}. However, in general
these are two different classes of equations.

To proceed further with equation (\ref{2:gl:Cond}), we need the concept of
conditional invariance which we recall for the convenience of the reader.

\noindent {\bf Definition.} \cite [Section 5.7]{Fush93} A PDE of
the form \BEQ \label{2:gl:c1} S\left(\vec{x},u,\underset{1}{u},...
\underset{m_1}{u}\right) =0 \EEQ (here $\underset{k}{u}$ is the
totality of $k^{\rm th}$-order derivatives) is {\it conditionally
invariant} under the operator
\BEQ \label{2:gl:Q}
Q = \xi^a (x_1,...,x_N, u)\partial_a + \eta(x_1,...,x_N,u)\partial_u
\EEQ
where $\eta$ and $\xi^a$ with $a=1,\ldots,N$ are smooth functions,
if it is invariant (in Lie's sense) under this operator only
together with an additional condition of the form
\BEQ
\label{2:gl:c2} S_Q\left(\vec{x},u,\underset{1}{u},\ldots
\underset{m_2}{u}\right) =0
\EEQ
that is, the overdetermined system
eqs.~(\ref{2:gl:c1},\ref{2:gl:c2}) is invariant under a Lie group
generated by the operator $Q$.

If the additional condition (\ref{2:gl:c2}) coincides with the
equation $Qu = 0$ then a {\it $Q$-conditional symmetry} is
obtained. The notion of $Q$-conditional symmetry coincides with
the notion of the non-classical symmetry as introduced in
\cite{Blum69} and the side condition as introduced in
\cite{Olve87}. The  notion  of the non-classical symmetry was
further developed in \cite{Levi89} where it is called `conditional
symmetry', these authors also considered linear second-order
partial differential equations as an additional condition for the
generation of conditional symmetries \cite[p. 2922]{Levi89}.

Following this definition, we impose the additional condition
\BEQ \label{2:gl:CS1}
W_N^{II} = 0
\EEQ
and then eq.~(\ref{2:gl:Cond}) reduces to the expression
\BEQ \label{2:gl:CS2}
A(t)^{1-z-N} W^I = f\left( A(t) u,\underset{1}{u}, A(t)^{-1}
\underset{11}{u}\right)
\EEQ
Clearly, having (\ref{2:gl:CS2}) and the condition
(\ref{2:gl:CS1}) we recover again eq.~(\ref{2:gl:Th2}).
Eq.~(\ref{2:gl:CS1}) is a multidimensional extension of the
well-known two-dimensional Monge-Amp\`ere equation, see \cite
{Guti01, Pogo75, Schu90}
\BEQ \label{2:gl:MA}
u_{11} u_{22} - u_{12}^2 = 0
\EEQ
The MIA of the $N$-dimensional Monge-Amp\`ere
equation (\ref{2:gl:CS1}), which we denote by $\mathfrak{M}_N$, is
well-known.

\noindent {\bf Lemma.} \cite[Theorem 1.10.1]{Fush93} {\it Consider
the $N$-dimensional Monge-Amp\`ere equation (\ref{2:gl:CS1}) where
$W_N^{II}$ is defined in (\ref{2:gl:wIII}) and also write $x_0
:=u$. Then a basis of the MIA $\mathfrak{M}_N$ is given by the
following operators}
\BEQ
\partial_i \;\; , \;\; x_i\partial_j \;\; , \;\; x_i x_j\partial_j
\EEQ
{\it where $i,j=0,1,\ldots,N$.}

It is easily seen that $\mathfrak{B}_N(z)$ and $\mathfrak{A}_N(z)$ are
subalgebras of $\mathfrak{M}_N$ because the `time' $t$ is only a parameter in
the Monge-Amp\`ere equation (\ref{2:gl:CS1}).
We therefore have our main result.

\noindent {\bf Theorem 3.} {\it A partial differential equation of
the form (\ref{2:gl:PDE}) is conditionally invariant under the
infinite-dimensional Lie algebra $\mathfrak{A}_N(z)$ if and only
if (i) it is of the form (\ref{2:gl:Th2}) and (ii) the
$N$-dimensional Monge-Amp\`ere equation (\ref{2:gl:CS1}) holds.}

We point out that we deal here with a non-Lie invariance because
the classical criterion of Lie invariance (see, e.g.,
\cite{Blum74}, \cite{Olve86}) does not admit any side condition.
Also, it is not the non-classical invariance in the Bluman-Cole
sense \cite{Blum69} because that kind of invariance admits only an
additional condition in the form of a {\it quasilinear first-oder}
PDE, see (\ref{2:gl:Q}), while the additional condition
Eq.~(\ref{2:gl:CS1}) is a {\it strongly non-linear second-oder}
PDE. Thus, we have found here a non-trivial example of a purely
conditional invariance as introduced by Fushchych and
collaborators \cite{Fush88,Fush93}. Of course, this could still
turn out to be a Q-conditional symmetry with a higher-order
symmetry operator, see e.g. \cite{Foka94,Zhda95}, and we hope to
study this elsewhere.

\noindent {\bf Remark 3.} One can easily check that Theorems 2 and
3 are still valid if one replaces the series of the operators
$Y_m^{(a)}$ arising in the algebras $\mathfrak{B}_N(z)$ and
$\mathfrak{A}_N(z)$ by the operators \BEQ \label{2:gl:R3}
Y_{\varphi}^{(a)}=\varphi_a(t)\partial_a \;\; ; \;\; a=1,\ldots,N
\EEQ where $\varphi_a(t)$ are arbitrary smooth functions and there
is no summation over the index $a$ here.

\section{Exact solutions of the conditionally invariant equation
(\ref{1:gl:inveq})}

It is easily seen that Eq.~(\ref{2:gl:Th2}) is a nonlinear PDE for
any fixed function $g$. Even the simplest examples of this class,
such as $W^I=0$ and $W^I=u^{1-z-N}$ which were studied in
\cite{Fush85,Cher85} are still non-linear equations. One possible
point of view is to consider the determinant $W^I$ (see
Eq.~(\ref{2:gl:wI})) as a generalization of the `velocity' $u_t$.
In the context of such a physical interpretation it appears to be
reasonable to study a generalized diffusion equation of the form
\BEQ \label{3:gl:gl}
W^I = u^{2-z-N} \Delta u + \gamma u^{1-z-N} u_a u_a
\EEQ
(where $\gamma$ is a constant) since it is the
simplest example of (\ref{2:gl:Th2}) containing the Laplacian
$\Delta=\partial_a\partial_a$. If we set $\gamma=2-z-N$,
Eq.~(\ref{3:gl:gl}) takes the form
\BEQ \label{3:gl:dgl}
W^I = \frac{\partial}{\partial x_1}\left( u^{2-z-N} u_1\right) + \cdots
+ \frac{\partial}{\partial x_N}\left( u^{2-z-N} u_N\right)
\EEQ
which for $N=2$ coincides with (\ref{1:gl:inveq}). In this section
we shall be looking for exact solutions of this equation, under
the condition $W_N^{II}=0$.

The one-dimensional case $N=1$ is trivial. The condition
$W_N^{II}=0$ reduces to $u_{xx}=0$ which implies $u=p(t)x+q(t)$.
Substituting this into (\ref{3:gl:dgl}) we arrive at the three
different cases $z=0$, $z=1$ and $z\ne 0,1$. The corresponding solutions
are $u=c xe^{t}+q(t)$, $u=cx+q(t)$ and $u=q(t)$,
respectively, where $q(t)$ remains arbitrary and $c$ is an arbitrary constant.

The first non-trivial case arises in two dimensions $N=2$. The
side condition (\ref{2:gl:CS1}) becomes the well-known
two-dimensional Monge-Amp\`ere equation. We want to use the
conditional symmetry obtained in section 2 to find exact solutions
to the pair (\ref{1:gl:inveq},\ref{1:gl:inveqMA}). According to
Theorem~3 and using standard techniques, we can seek for exact
solutions of (\ref{1:gl:inveq}) by solving the associated Lagrange
system which reads
\BEQ \label{3:gl:Lag}
\frac{\D t}{e_1 z t^{n+1}} =
\frac{\D x_1}{e_1(n+1)t^n x_1 +e_2t^{m+1/z}} =
\frac{\D x_2}{e_1(n+1)t^n x_2 + e_3 t^{k+1/z}} =
\frac{\D u}{e_1(n+1)t^n u}
\EEQ
where $e_1,e_2$ and $e_3$ are arbitrary parameters. In
general, we might have used in (\ref{3:gl:Lag}) finite series,
viz. $\sum_{i=n_1}^{n_2} e_{1i}(i+1) t^i$, $\sum_{i=m_1}^{m_2}
e_{2i} t^{i+1/z}$ and $\sum_{i=k_1}^{k_2} e_{3i} t^{i+1/z}$
instead of $e_1 t^{n+1}, e_2 t^{m+1/z}, e_3 t^{k+1/z}$,
respectively (where $n_1<n_2 \in \mathbb{Z}$ and $m_1<m_2, k_1<k_2
\in \mathbb{Z}-1/z$ and the $e_{ji}$ may be taken to be real).
However,  our aim is to show that even in the simplest case
presented by (\ref{3:gl:Lag}) the conditional symmetry leads to
non-trivial solutions. Indeed, setting $e_2=e_3=0$ (the parameter
$e_1\ne 0$ since it corresponds to the conditional symmetry
generator $X_n$) and solving the resulting linear system
(\ref{3:gl:Lag}), we obtain the ansatz
\BEA u &=& t^{(n+1)/z}
\vph\left(\omega_1, \omega_2\right) \label{3:gl:ansatz1} \\
\omega_a &=& x_a t^{-(n+1)/z} \;\;\;\;\; ; \;\; a=1,2 \nonumber
\EEA
if $z\ne 0$ and
\BEQ \label{3:gl:ansatz2}
u = x_1\vph(\omega,t) \;\; , \;\; \omega=\frac{x_1}{x_2}
\EEQ
if $z=0$.

Consider the case $z\ne 0$. Substituting the ansatz (\ref{3:gl:ansatz1}) into
eqs.~(\ref{1:gl:inveq},\ref{1:gl:inveqMA}), we arrive at the system
\BEA
\vph
\left( \frac{\partial^2 \vph}{\partial\omega_1^2}+
\frac{\partial^2 \vph}{\partial\omega_2^2}\right) &=& z
\left(\frac{\partial\vph}{\partial\omega_1}\right)^2 + z
\left(\frac{\partial\vph}{\partial\omega_2}\right)^2 \nonumber
\\
\frac{\partial^2 \vph}{\partial\omega_1^2} \frac{\partial^2
\vph}{\partial\omega_2^2}  &=& \left(\frac{\partial^2
\vph}{\partial\omega_1\partial\omega_2}\right)^2 \label{3:gl:2sys}
\EEA
It turns out that the general solution of this system can be
written down explicitly. Indeed, the local substitution
\BEQ \label{3:gl:subs}
\vph = \left\{ \begin{array}{ll} \exp \phi & \mbox{\rm ~~;~ $z=1$} \\
                            \phi^{1/(1-z)} & \mbox{\rm ~~;~ $z\ne 1$}
\end{array} \right.
\EEQ
with $\phi=\phi(\omega_1,\omega_2)$ reduces the first equation of
(\ref{3:gl:2sys}) to the Laplace equation $\Delta\phi=0$ with the
general solution
\BEQ
\phi= f(\omega_1 +\II\omega_2) + g(\omega_1 -\II\omega_2)
\EEQ
where $f$ and $g$ are arbitrary smooth functions and $\II^2=-1$.
When reinserting this into the second equation (\ref{3:gl:2sys}), we obtain an
equation which can be separated into two ordinary differential
equations for the functions $f$ and $g$ which are elementarily integrable. We
merely give the final result. For $z=1$ we find
\BEQ
u = \left( c_1(x_1+\II x_2)+d_1 t^{n+1}\right)^{\alpha}
\left( c_2(x_1-\II x_2)+d_2 t^{n+1}\right)^{1-\alpha}
\EEQ
where $\alpha$ is constant. For $z\ne 1$ the result is
\BEQ
u=\left[ \left( c_1(x_1+\II x_2)+d_1 t^{n+1}\right)^{1-z}+
\left( c_2(x_1-\II x_2)+d_2 t^{n+1}\right)^{1-z}\right]^{1/(1-z)}
\EEQ
and where $c_{1,2}$ and $d_{1,2}$ are arbitrary integration constants.

In general, these solutions are complex. In order to make them real, we must
have $c_1=c_2=c^2>0$, $d_1=d_2^*=c^2(e_1+\II e_2)$ and $\alpha=\smfrac{1}{2}$,
where $c>0$ and $e_{1,2}\in\mathbb{R}$. Under these restrictions, the above
solutions become
\BEQ \label{lsg1}
u = c \sqrt{ \left(x_1+e_1t^{n+1}\right)^2 + \left(x_2+e_2t^{n+1}\right)^2\;}
\EEQ
for $z=1$ and
\BEA
u &=& c\left[
\left[ \left(x_1+e_1 t^{(n+1)/z}\right)
+\II \left(x_2+e_2 t^{(n+1)/z}\right) \right]^{1-z}
+\left[ \left(x_1+e_1 t^{(n+1)/z}\right)
-\II \left(x_2+e_2 t^{(n+1)/z}\right) \right]^{1-z}
\right]^{1/(1-z)}
\nonumber \\
&=& 2c \sqrt{\left(x_1+e_1 t^{(n+1)/z}\right)^2 + \left(x_2+e_2
t^{(n+1)/z}\right)^2\;} \left(\cos\left[(1-z)\arctan \frac{x_2+e_2
t^{(n+1)/z}}{x_1+e_1 t^{(n+1)/z}}\right]\right)^{1/(1-z)}
\label{lsg2} \EEA
for $z\ne 1$, respectively. From the second line
in Eq.~(\ref{lsg2}) the reality of the solution is explicit and we
also see that the solution (\ref{lsg1}) obtained for $z=1$ can be
recovered by taking the limit $z\to 1$ in (\ref{lsg2}). Therefore,
Eq.~(\ref{lsg2}) gives the general solution of the system
(\ref{3:gl:2sys}) and we have obtained a four-parameter family of
functions, depending on `time' and on two `space' coordinates,
which simultaneously solve the generalized diffusion equation
(\ref{1:gl:inveq}) and the Monge-Amp\`ere equation
(\ref{2:gl:MA}). Of these four parameters, $c$ merely is a trivial
normalization constant and $e_{1,2}$ control the scaling between
`time' and `space' directions. The functional form of the
solutions is controlled by the dynamical exponent $z$. Finally,
the integer parameter $n$ selects the rescaling between `time' and
`space' coordinates.

Different exact solutions of (\ref{1:gl:inveq}) which are not a
solution of a Monge-Amp\`ere equation might be found by using only
the Lie symmetries of the algebra $\mathfrak{B}_N(z)$ obtained in
Theorem 2 but we shall not go into this here.

Consider now Eq.~(\ref{1:gl:inveq}) for $z=0$, when it becomes
\BEQ \label{3:13}
\det\left[\begin{array}{ccc} u_t    & u_1    & u_2 \\
                             u_{t1} & u_{11} & u_{12} \\
                             u_{t2} & u_{21} & u_{22} \end{array}\right]
= \Delta u, \quad u=u(t,x_1,x_2).
\EEQ
Substituting the ansatz
(\ref{3:gl:ansatz2}) into Eq.~(\ref{3:13}), we arrive at the equation
\BEQ \label{3:14}
-\omega^2\vph \frac{\partial
\vph}{\partial t}\left(\omega \frac{\partial^2
\vph}{\partial\omega^2}+2\frac{\partial
\vph}{\partial\omega}\right)=(1+\omega^2)\left(\omega
\frac{\partial^2 \vph}{\partial\omega^2}+2\frac{\partial
\vph}{\partial\omega}\right). \EEQ
Simultaneously we substitute
one into condition (\ref{2:gl:MA}) and see that it simply leads to
the identity. Since Eq.~(\ref{3:14}) decomposes into two
independent equations
\BEQ
-\omega^2\vph \frac{\partial \vph}{\partial t} = 1+\omega^2 \;\; , \;\;
\omega \frac{\partial^2\vph}{\partial\omega^2}+ 2\frac{\partial
\vph}{\partial\omega}=0
\EEQ
we easily obtain its general
solution. Inserting into (\ref{3:gl:ansatz2}), we arrive at two
families of exact solutions of Eq.~(\ref{3:13})
\BEA
u &=& \sqrt{ 2t(x_1^2+x_2^2)+\psi(x_1/x_2)x_1^2}\label{3:15} \\
u &=& x_1\psi_1(t)+x_2\psi_2(t) \label{3:16}
\EEA
where $\psi,\psi_1,\psi_2$ are arbitrary smooth functions.

Finally, we point out how Remark~3 can be used for generalization
of the exact solutions obtained. Operators (\ref{2:gl:R3})  with
$N=2$ generate the invariance transformations
\BEQ \label{3:17}
t'=t\;\; , \;\;
x_1'=x_1+e_1\vph_1(t) \;\; , \;\;
x_2'=x_2+e_2\vph_2(t) \;\; , \;\;
u'=u
\EEQ
where $e_1$ and $e_2$
are real parameters. Hence, we can transform any known solution
$u=u^0(t,x_1,x_2)$ of Eq.~(\ref{1:gl:inveq}) into a new solution
of the form $u=u^0(t,x_1+e_1 \vph_1(t),x_2+e_2 \vph_2(t))$ using
formul{\ae} (\ref{3:17}). In the particular case, the solution
(\ref{lsg2}) can be generalized to the form
\BEQ \label{3:18}
u= 2c \sqrt{\left(x_1+e_1 \vph_1(t)\right)^2 + \left(x_2+e_2
\vph_2(t)\right)^2\;} \left(\cos\left[(1-z)\arctan \frac{x_2+e_2
\vph_2(t)}{x_1+e_1 \vph_1(t)}\right]\right)^{1/(1-z)}.
\EEQ
By an appropriate choice of two arbitrary functions $\vph_1(t)$ and
$\vph_2(t)$ in the solution (\ref{3:18}), one can satisfy a wide
range of boundary conditions that can be given for
Eq.~(\ref{1:gl:inveq}).

\noindent {\bf Remark 4.} A long list of exact solutions of the
Monge-Amp\`ere equation is given in \cite[Section 1.10]{Fush93}.
However, the solutions (\ref{3:15}) and  (\ref{3:18}) are not
contained in this list and appear to be new. Considering the
'time' $t$ only  as a parameter and taking into account the
space-translation invariance of the  Monge-Amp\`ere equation,
these solutions can be united to the form $u= x_1\vph(x_1/x_2)$.
Moreover, one can check that the obvious generalization of this, namely
\BEQ
u= x_1\vph\left(\frac{x_1}{x_2},...,\frac{x_1}{x_N} \right)
\EEQ
where $\vph$ is an arbitrary function of $N-1$
arguments, is a new solution of the $N$-dimensional Monge-Amp\`ere
equation (\ref{2:gl:CS1}).

\section{Conclusions}

In this paper, we have given in Theorems~2 and 3 a complete
description of first- and second-order PDEs of the form
(\ref{2:gl:PDE}) which are (conditionally) invariant under the
infinite-dimensional Lie algebras $\mathfrak{A}_N(z)$ and
$\mathfrak{B}_N(z)$. While there is a
wide class of PDEs (\ref{2:gl:Th2}) which are invariant under  Lie
algebras $\mathfrak{B}_N(z)$, there is no equations with the
$\mathfrak{A}_N(z)$ Lie symmetry. However, we have shown that the
same class (\ref{2:gl:Th2}) contains all possible PDEs which
which are {\em conditionally} invariant under the Lie
algebra $\mathfrak{A}_N(z)$ and that the well-known Monge-Amp\`ere
equation is the required additional condition.
Both theorems can be extended on the case of the Lie algebras
$<\mathfrak{A}_N(z), J_{ab}>$ and $<\mathfrak{B}_N(z), J_{ab}>$,
where the operators $J_{ab}$ are given in (\ref{1:gl:XYJ}).
We applied these general results to the non-linear diffusion equation
with the generalized time-derivative (\ref{1:gl:inveq}). Using the conditional
invariance, some families of exact solutions
of this strongly non-linear equation
have been found. Simultaneously, a new exact solution of
the two-dimensional Monge-Amp\`ere equation was obtained.

In a wider perspective, we note that the result obtained here cannot be
directly extended on the case of the  Lie algebras
$\mathfrak{A}_N(z)$ and $\mathfrak{B}_N(z)$
with `mass' operators because of the
well-known fact that classes of PDEs invariant under the Galilei
algebra with the `mass' operator and the massless  Galilei
algebra have essentially different structures. We hope to be able to
return to this open problem elsewhere.

\medskip

\centerline {\bf Acknowledgments}

\medskip

R.Ch. was supported by the CNRS (D\'ecision N$^{o}$ 2176 from the
29$^{\rm th}$ of September 2003). The authors are grateful to the
referee for pointing out the reference \cite{Levi89}.



\begin{thebibliography}{999}

\bibitem{Blum69} G.W. Bluman and J.D. Cole,
The general similarity solution of the heat equation.
J. Math. Mech. {\bf 18} (1969),\, 1025-1042.
\bibitem{Blum74} G.W. Bluman  and J.D. Cole,
{\it Similarity Methods for Differential Equations.}
Springer, Heidelberg, 1974.
\bibitem{Bren03} Y. Brenier, U. Frisch, M. H\'enon, G. Loeper, S. Matarrese,
R. Mohayaee and A. Sobolskii, Reconstruction of the early universe
as a convex optimization problem, Mon. Not. Astron. Soc. {\bf 346} (2003),
501--524 .
\bibitem{Caff99} L.A. Caffarelli, in M. Christ et {\it al.} (eds), {\it
Harmonic analysis and partial differential equations}, University
of Chicago Press (London 1999), 117--126.
\bibitem{Cher85} R. Cherniha,
Two-dimensional nonlinear equations that are invariant under the
Galilei algebra, in: Group-theoretical Investigations of Equations
of Mathematical Physics,  Institute of Mathematics, Ukrainian
Acad.Sci., Kyiv, (1985)\, 107--114, in russian.
\bibitem{Cher98} R. Cherniha and M. Serov,
Symmetries, Ans\"atze  and Exact Solutions of  Nonlinear
Second-order Evolution Equations with Convection Term,
Euro.\ J.\ Appl.\ Math.\ {\bf 9} (1998),\, 527--542.
\bibitem{Cher01} R. Cherniha,
Nonlinear Galilei-invariant PDEs
with infinite-dimensional Lie symmetry, J. Math. Anal. Appl.
{\bf 253} (2001), 126-141 .
\bibitem{Erco03} N. Ercolani, R. Indik, A.C. Newell and T. Passot,
Global description of patterns far from onset: a case study,
Physica {\bf D184} (2003), 127--140.
\bibitem{Foka94} A.S. Fokas and Q.M. Liu, Nonlinear interaction of
traveling waves of nonintegrable equations,
Phys. Rev. Lett. {\bf 72} (1994), 3293-3296.
\bibitem{Fush85} W.I. Fushchych and R. Cherniha, The Galilean
relativistic principle and nonlinear partial differential
equations, J. Phys. A Math. Gen. {\bf 18} (1985), 3491--3503.
\bibitem{Fush88} W.I. Fushchych, M.I. Serov and W. Chopyk,
Conditional invariance and nonlinear heat equations, Dopovidi
Akad.Nauk Ukrainy, Ser. A
(Proc. Ukrainian Acad. Sci., Ser. A) {\bf N 9} (1988)\, 17--21, in russian.
\bibitem{Fush93} W.I. Fushchych, W.M. Shtelen and M.I. Serov,
{\it Symmetry analysis and exact solutions of equations of
nonlinear mathematical physics}, Kluwer, Dordrecht 1993.
\bibitem{Guti01} C.E. Guti\'errez, {\it The Monge-Amp\`ere equation},
Birkh\"auser, Basel 2001.
\bibitem{Henk94} M. Henkel, Schr\"odinger Invariance and Strongly
Anisotropic Critical Systems, J. Stat. Phys. {\bf 75} (1994), 1023--1061.
\bibitem{Henk02} M. Henkel, Phenomenology of local scale-invariance: from
conformal invariance to dynamical scaling, Nucl. Phys.
{\bf B641} (2002), 405--486 .
\bibitem{Henk03} M. Henkel and J. Unterberger, Schr\"odinger-invariance
and space-time symmetries, Nucl. Phys. {\bf B660} (2003), 407--435.
\bibitem{Ivas97} E.V. Ivashkevich, Symmetries of the stochastic Burgers
equation, J. Phys. A Math. Gen. {\bf 30} (1997), L525--L533.
\bibitem{Levi89} D. Levi and P. Winternitz, Non-classical symmetry reduction:
example of the Boussinesq equation, J. Phys. A Math. Gen. {\bf 22} (1989),
2915-2924.
\bibitem{Newe96} A.C. Newell, T. Passot, C. Bowman, N. Ercolani and R. Indik,
Defects are weak and self-dual solutions of the Cross-Newell phase
diffusion equation for natural patterns, Physica {\bf D97} (1996), 185--205.
\bibitem{Nied72} U. Niederer, The maximal kinematical invariance group of
the free Schr\"odinger equation, Helv. Phys. Acta. {\bf 45}, 802--810 (1972).
\bibitem {Olve86} P. Olver,
{\it Applications of Lie Groups to Differential Equations.}
Springer, Heidelberg, 1986.
\bibitem{Olve87} P. Olver and P. Rosenau, Group-invariant solutions of 
differential equations, SIAM J.~Appl. Math.~{\bf 47} (1987),~263--278.
\bibitem{Pogo75} A.V. Pogorelov, {\it Multidimensional Minkowski
problem}, Nauka, Moscow, 1975, in russian.
\bibitem{Schu90} F. Schulz, {\it Regularity theory for quasilinear elliptic
systems and Monge-Amp\`ere equations in two dimensions}, Springer
Lecture Notes in Mathematics vol. 1445, Springer, Heidelberg 1990.
\bibitem{Volo99} A. Volovich, Domain walls in MQCD and Monge-Ampere
equation, Phys. Rev. {\bf D59} (1999), 065005.
\bibitem{Zhda95} R. Zhdanov, Conditional Lie-B\"acklund symmetry and reduction
of evolution equations, J. Phys. A Math. Gen. {\bf 28} (1995), 3841-3850.
\end{thebibliography}
\end{document}